# Fast computation of scattered fields with arbitrary beam scattering

Yuval Kashter[a,*], Eran Falek[b], Pavel Ginzburg[a]

[a] Department of Electrical Engineering, Tel-Aviv university, Ramat Aviv, Tel Aviv 69978, Israel. E-mail: yuvalkashter@mail.tau.ac.il

[b] School of Electrical and Computer Engineering, Ben-Gurion university, Beer-Sheva 8410501, Israel.

**Abstract:**

Electromagnetic scattering on a sphere is one of the most fundamental problems, which has a closed form analytical solution in the form of Mie series. Being initially formulated for a plane incident wave, the formalism can be extended to more complex forms of incident illumination. Here we present a fast calculation approach to address the scattering problem in the case of arbitrary illumination, incident on a spherical scatterer. This method is based on the plane wave decomposition of the incident illumination and weighted integration of Mie solutions, rotated to a global coordinate system. Tabulated solutions, sampled with an accurately level of sparsity, and efficient rotational transformations allow performing fast calculations on electrically large structures, outperforming capabilities relatively to other numerical techniques. Our approach is suitable for real-time analysis of electromagnetic scattering from electrically large objects, which is essential for monitoring and control applications, such as optomechanical manipulation, scanning microscope, and fast optimization algorithms.

*corresponding author: Yuval Kashter

## 1. Introduction

Electromagnetic scattering is one of the most important physical phenomena, related to many practical applications [1]. For instance, optical imaging systems typically based on efficient collection and analysis of scattered light waves. Optomechanical manipulation is another example, where light induces mechanical forces on particles via scattering and absorption [2],[3].

Comprehensive scattering theory was first introduced by Gustav Mie in 1908, who analyzed the interaction of spherical particles with plane incident waves [1,4]. Since then, this formalism has advanced to more complicated and practical cases (e.g. particles in an absorbing medium [5], aspherical scatterers [6], anisotropic particles [7], [8],[9],[10], coated spheres [11], and [12] for a recent perspective review) and nowadays it is routinely used for analysis of numerous Nano photonic scenarios, including optical magnetism [13], Huygens metasurfaces [14], and other frontier optical devices (e.g. [15] for a review).

An advanced form of Mie scattering named the generalized Lorentz-Mie theory (GLMT) [16], [17] is based on the method of separation of variables in a modified coordinate system. Consequently, a linear system that satisfy the Helmholtz equation can be solved by the above-mentioned modified coordinate system. More recently, a different approach was presented by Cui and Han [18]. According to their study, combination of two advanced computation methods, namely method of moments (MoM) [19] and finite element method (FEM) [20], provides a generalized rigorous solution for many cases of arbitrary shaped particles and illumination sources. The above-mentioned techniques provide solutions for scattered fields, e.g. in the case of Gaussian beam interaction with a spheroid. However, for every particular phase pattern and particle position, a relatively complicated calculation process should be performed, which might require significant computational efforts. Therefore, applications that require real-time calculations, such as particle tracking, fast changing illumination field, and others require development of new approaches.

Here, we present a different approach to perform fast and efficient calculations of scattered fields, resulted from arbitrary incident phase patterns that illuminate a spherical particle of any size (not necessarily electrically small). In this method, we generate a database of scattering solutions. Each element in the data base corresponds to a scattering of a plane wave, incoming at a certain angle and polarization. Hence, instead of calculating a general solution for each case of an arbitrary phase pattern, we perform a plane wave decomposition of the incoming wave front and retrieve the multiple Mie solutions (from the above mentioned database) to each one of the spatial Fourier components. At the next step, all the contributions are summed up (integrated) with a careful consideration of amplitude, phase, polarization, and propagation direction in the initial plane wave decomposition.

The challenge in the proposed technique is that each tilted plane wave and its corresponding Mie solution is represented in a different rotated coordinate system. In order to apply the plane wave decomposition technique, the solutions in the database should be represented at the same global coordinates system. Mie solutions are typically given in spherical coordinates, hence a set of rotations with non-uniform sampling density should be applied.

In the case of electrically small particles, scattered fields are described by lower order multipoles and are rather smooth, simplifying obtaining a high numerical accuracy with moderately low computational effort. However, scattered field patterns of electrically large particles introduce fast oscillations in space, requiring accurate and dense numerical sampling. This problem might be severe in cases where the coordinate system is rotated in different angles. However, the representations of the rotated coordinate system in the global system results in nonlinear changes in the axes incremental distributions.

In previous Mie studies, choosing a sufficient number of multipole expansion terms, with respect to the size of the particle and the ratio between the refractive indexes, was thoroughly discussed (e.g., [21]). Nevertheless, estimating the effect on the overall numerical error due to a limited number of samples (along the spherical coordinate system) has not been accurately studied and will be shown here to be important in non-paraxial cases.

The manuscript is organized as follows: section 2 is a brief review of Mie scattering [3] in light of consideration of angular sampling densities and particle sizes. Section 3 presents a method to calculate the scattered field resulting from a rotated plane wave in terms of a global system. In section 4, the plane wave decomposition based scattering calculation of arbitrary phase distribution is presented, including analytical overview and simulation results of a special case of a focused Gaussian beam.

**2. Mie scattering of a single incident plane wave**

The Mie solution is given in the form of multipole series. Each term in this expression is described by the vector spherical harmonics that satisfy Helmholtz equation. As the scatterer gets larger relative to the wavelength, and as its refractive index increases (relative to the surrounding medium), more terms in the multipole expansion are required for an accurate electromagnetic description.

Higher multipoles have fast oscillating features at the far field. In other words, scattering problems, involving electrically large particles, require a denser numerical sampling of the far field. Sparse sampling density, however, might cause an accumulative numerical error in estimating different types of observables (e.g. total electromagnetic field, electromagnetic forces). The subject of numerical sampling density is particularly important when the incident field has a non-uniform wave front with fast-oscillating features. Consequently, exploring the effect of the sampling sparsity on a range of possible electromagnetic scenarios is valuable.

In this section, we introduce an error estimation method to address the influence of the sampling density. For this reason, a standard Mie solution will be analyzed. Fig. 1 presents the general layout of a scattering problem - a homogenous nonmagnetic spherical particle with radius $a$ and refractive index $n_p$, surrounded by a uniform medium with refractive index $n_m$. The origin of the global Cartesian coordinate system $\boldsymbol{r_0} = [x, y, z]$ is located at the center of the scatterer. The linearly polarized incident plane wave propagates along $\hat{\boldsymbol{k}}_i$, which is parallel to $z$ axis. The resulted electric field and magnetic induction field, denoted by $E_i$ and $B_i$, are oriented along *x* and *y* axes, respectively. The orthogonal unit vectors $\hat{u}_\theta, \hat{u}_\phi$ and $\hat{u}_r$ represent a spherical coordinate system, in which Mie solutions are presented; $\hat{k}_s$ represents the propagation direction of the scattered field in every direction. The scattered electric and magnetic induction fields are denoted $E_s$ and $B_s$, respectively.

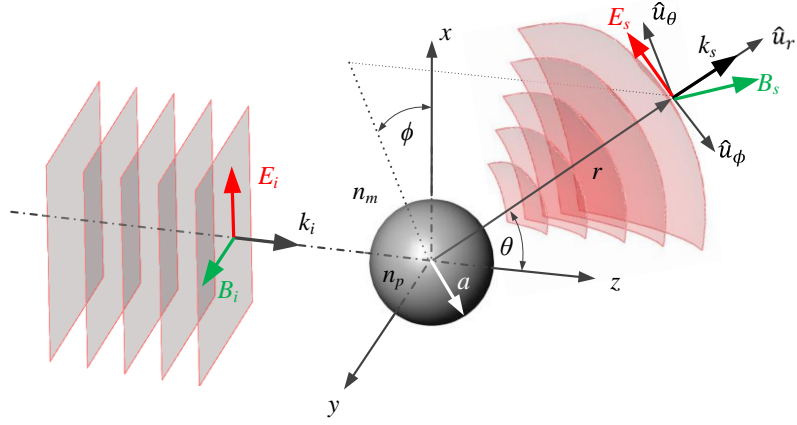

Fig. 1. Geometry of the basic scattering problem - a spherical particle is illuminated by a plane wave.

The monochromatic incident field phasors (time dependency is $e^{-i\omega t}$) are given by:

$$\begin{aligned} \boldsymbol{E}_i(\boldsymbol{k}_i) &= E_i e^{i\boldsymbol{k}_i \cdot \boldsymbol{r}} \hat{x} \\ \boldsymbol{B}_i(\boldsymbol{k}_i) &= \frac{n_m E_i}{c} e^{i\boldsymbol{k}_i \cdot \boldsymbol{r}} \hat{y} \end{aligned} \quad (1)$$

where, $\boldsymbol{k}_i = \frac{2\pi}{\lambda_0} n_m \hat{z}$, $c$ is the speed of light in vacuum and $\lambda_0$ is the wavelength in vacuum.

Scattered fields are given by:

$$\begin{aligned} \boldsymbol{E}_s(r, \hat{\boldsymbol{k}}_s; \hat{\boldsymbol{k}}_i) &= E_i \boldsymbol{f}(\hat{\boldsymbol{k}}_s, \hat{\boldsymbol{k}}_i) \frac{e^{ik_m r}}{r} \\ \boldsymbol{B}_s(r, \hat{\boldsymbol{k}}_s; \hat{\boldsymbol{k}}_i) &= \frac{n_m E_i}{c} \hat{\boldsymbol{k}}_s \times \boldsymbol{f}(\hat{\boldsymbol{k}}_s, \hat{\boldsymbol{k}}_i) \frac{e^{ik_m r}}{r} \end{aligned} \quad (2)$$

where $k_m = \frac{2\pi}{\lambda_0} n_m$ and $r \gg \lambda_0/n_m$ is the distance from the center of the sphere. The term $\boldsymbol{f}(\hat{\boldsymbol{k}}_s, \hat{\boldsymbol{k}}_i)$ is the scattering amplitude, relating the scattered and the incident fields as follows:

$$\boldsymbol{f}(\hat{\boldsymbol{k}}_s, \hat{\boldsymbol{k}}_i) = \frac{1}{-ik_m} \sum_n^N \frac{2n+1}{n(n+1)} \left[ (a_n \tilde{\pi}_n + b_n \tilde{\tau}_n) \cos\phi \hat{u}_\theta - (a_n \tilde{\tau}_n + b_n \tilde{\pi}_n) \sin\phi \hat{u}_\phi \right], \quad (3)$$

where the $a_n$ and $b_n$ are the Mie coefficients, the terms $\tilde{\pi}_n$ and $\tilde{\tau}_n$ are "angle-dependent functions" [1]. $\boldsymbol{f}(\hat{\boldsymbol{k}}_s, \hat{\boldsymbol{k}}_i)$ depends on ($\theta \in [0,\pi], \phi \in [0,2\pi]$), which should be sampled on a finite set of discrete coordinates. Dividing $\theta$ and $\phi$ into a large number of segments will increase the sampling density, and consequently, will decrease the accumulated numerical errors. However, in order to reduce computational efforts (especially relevant to real-time applications), the sampling density should be reduced and, under certain circumstances, can affect the results. Hereafter, we will provide an assessment of the required sampling density, investigating the conversance of the total scattered power.

The scattering cross-section ($\sigma_{scat}$) can be calculated in two ways. The first one is given in a closed form by analytical integration of spherical harmonics as follows:

$$\sigma_{scat}^{(1)} = \frac{2\pi}{k_m^2} \sum_{n=1}^{N}(2n+1)(|a_n|^2 + |b_n|^2). \tag{4}$$

Differential cross-section, on the other hand, is directly derived from the scattering amplitude:

$$\frac{d\sigma_{scat}}{d\Omega} = \left|\boldsymbol{f}(\widehat{\boldsymbol{k}}_s, \widehat{\boldsymbol{k}}_i)\right|^2, \tag{5}$$

where $d\Omega$ represents a unit solid angle around $\widehat{k}_s$. Integrating the differential cross-section over all scattering angles will eventually provide the same expression of $\sigma_{scat}^{(2)}$:

$$\sigma_{scat}^{(2)} = \oint_\Omega \left|\boldsymbol{f}(\widehat{\boldsymbol{k}}_s, \widehat{\boldsymbol{k}}_i)\right|^2 d\Omega. \tag{6}$$

In terms of discrete representation, following Eq. 6, the scattering cross-section can be written in the form of:

$$\sigma_{scat}^{(2)} = \sum_{l_1=1}^{L_1} \sum_{l_2=1}^{L_2} \left|\boldsymbol{f}(\theta_{l_1}, \phi_{l_2})\right|^2 \sin(\theta_{l_1})\Delta\theta\Delta\phi, \tag{7}$$

where $L_1$ and $L_2$ are the number of discrete samples along $\theta$ and $\phi$, respectively; $\Delta\theta = \frac{\pi}{L_1}$ and $\Delta\phi = \frac{2\pi}{L_2}$ are the corresponding sampling densities. For the sake of simplicity, to get an equal sampling density along $\theta_{l_1}$ and $\phi_{l_2}$, the relation $L_2 = 2L_1 = L$ is chosen, where $L$ is the sampling density rate in our notations. From Eq. 4, one can conclude that regardless of the discretization effects, the scattering cross section $\sigma_{scat}^{(1)}$ is given only by the summation of Mie coefficients, as long as one is taking into consideration a sufficient number of multipoles. However, the numerical representation of the scattering cross-section $\sigma_{scat}^{(2)}$ (Eq. 7) is directly affected by the sampling density. Hence, comparing the numerical $\sigma^{(2)}{}_{scat}$ and exact $\sigma_{scat}^{(1)}$ expression provides an efficient numerical error estimation. The numerical error, derived from the scattering cross-section, is defined as follows:

$$\eta_{scat} = \frac{\left|\sigma_{scat}^{(1)} - \sigma_{scat}^{(2)}\right|}{\sigma_{scat}^{(1)}} \times 100\%. \tag{8}$$

Fig. 2 shows the results of analysis for several sphere's sizes (excitation wavelength $\lambda_0 = 633\ nm$, surrounding medium is water ($n_m$=1.33), glass particle $n_p$=1.5). Mie coefficients were calculated for a range of size parameters $k_m a \in [1,100]$, where number of multipoles (*N*) increases with respect to $k_m a$ (Fig.2 (a-c)). In Fig.2 (a-c), three examples of Mie coefficients are plotted, according to $k_m a = 1, 20,$ and 50, respectively. The corresponding Mie coefficients $a_n$ and $b_n$ are presented in Fig 2. (a-c), where the number of multiple expansions is N=5, 30 and 65, respectively. Negligible magnitudes of Mie coefficients in each case ($a_2$-$a_5$ and $b_2$-$b_5$, in Fig.2(a) ; $a_{22}$-$a_{30}$ and $b_{22}$-$b_{30}$ in Fig.2(b); $a_{55}$-$a_{65}$ and $b_{55}$-$b_{65}$ in Fig.2(c)), indicate that a sufficient number of multipoles has been chosen (relatively to $k_m a$ and $n_p$). Figures 2 (e-f) present the three corresponding scattering amplitudes, where the green curve represents the distribution in the *x-z* plane (E-plane), and the blue curve represents the distribution in the *y-z* lane (H-plane). Figs. 2(g-i) present the errors (Eq. 8) for three different sampling density rates (L=128, 256, and 512).

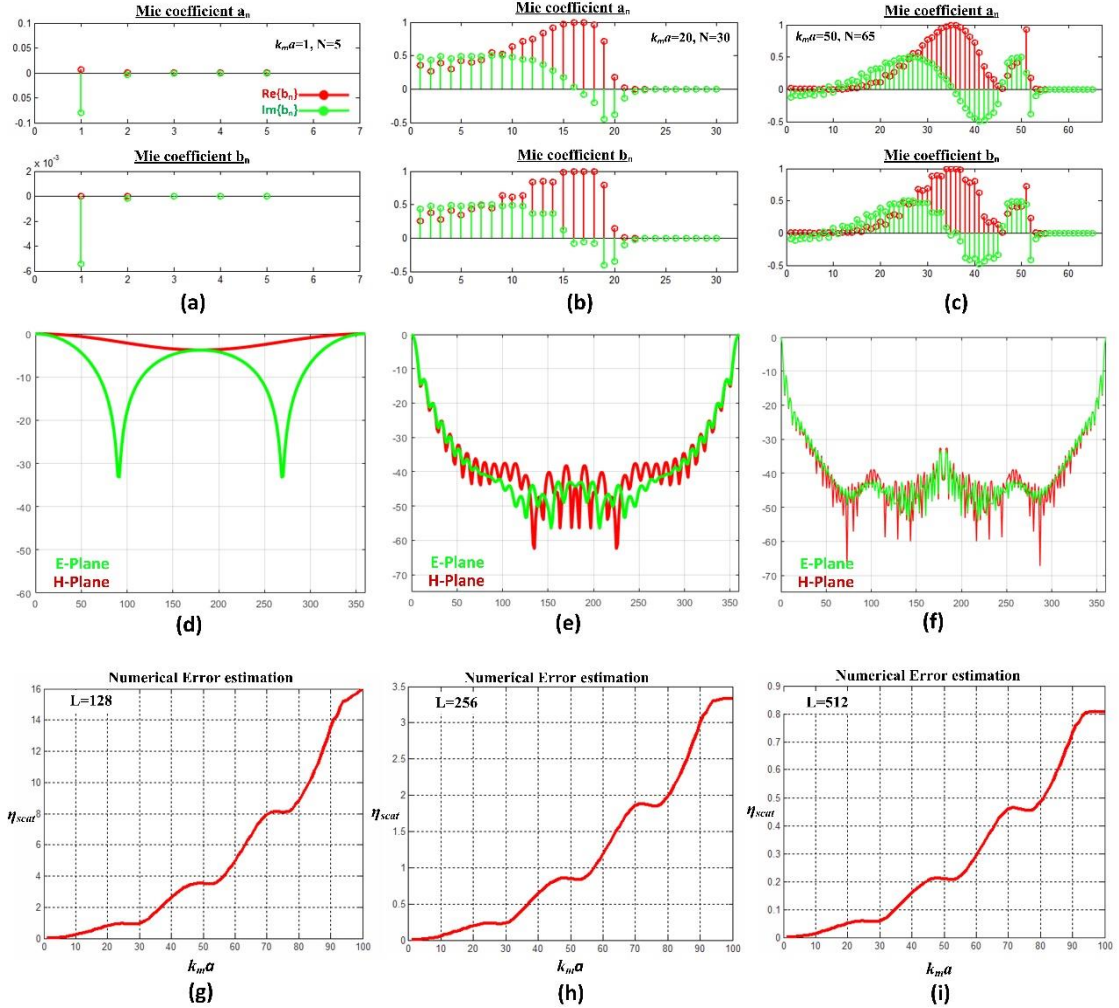

Fig. 2 Impact of angular sampling density on Mie scattering. (a-c) Mie coefficients $a_n$ and $b_n$ for three size parameters ($k_m a$=1, 20, and 50, respectively). Real and imaginary parts are represented with red and green colors, respectively; (d-f) Normalized scattering amplitudes at E and H planes, as the function of $\theta$ in a logarithmic scale. The E-plane and H-plane represent the scattering amplitudes in $\phi = 0$ and $\frac{\pi}{2}$, respectively; (g-i) Relative error in total scattering ($\eta_{scat}, Eq.8$) versus the size parameter for three sampling density rates $L$= 128, 256, and 512, respectively.

In the first example, presented in Fig. 2 (a,d), the size of the sphere's radius is 8.3 times smaller than the free space wavelength. As a result, the spatial variation of the scattering amplitude (Fig.2 (d)) is low, and therefore, the sampling density rate *L* can be smaller than 128, while the resulting numerical error $\eta_{scat}$ remains negligible. In the second case, presented in Fig.2 (b,e), the sphere radius is 2.4 times larger than the wavelength. In this case a significant spatial variation of the scattering amplitude can be noticed (Fig.2 (e)). According to Fig, 2(h), in order to decrease $\eta_{scat}$ to be lower than 0.5%, *L* should be larger than 256. In the third example, the radius is 11.96 times the wavelength, and therefore, the spatial variation is significantly larger than in previous cases (Fig.2 (i)). Consequently, here *L* is chosen to be 256 and the numerical error $\eta_{scat}$ will be almost 1%. Nevertheless, if one desires to decrease the error beneath 1%, the density rate should be chosen to be higher (e.g., according to Fig. 2, by choosing L=512 the resulting $\eta_{scat}$ is approximately 0.2%).

## 3. Rotation of Mie solution in a global coordinate system

An arbitrary illumination can be decomposed into a superposition of plane waves. A reliable calculation of the scattered field, in this case, requires adapting Mie solutions for incoming plane waves propagating in any incident angle and polarization direction. Weighted summation of these results will provide an accurate solution for the scattering problem.

In order to apply the beforehand discussed superposition principle, 'rotated' Mie solutions should be represented in terms of to the same 'global' coordinate system (taking into consideration the polarization state and the incident wavevector). In this case, the transformation should be performed using spherical harmonics-based expressions for the multipole decomposition.

Transformations of this kind have been already presented before, e.g. in a series of five manuscripts by G. Gouesbet et. al. [22], [23], [24], [25], and [26]. Specifically, the tilted incidence is treated in [25]. This approach considers the plane wave as a special case of on-axis beams with cylindrical symmetry. The beam shape coefficients (used in the above mentioned GLMT studies [16], [17]) is then modified by the tilting angles. As a result, the mathematical expression for the transformation is complicated and requires additional calculations. For example, in order to modify the above mentioned angle-dependent functions (presented if Eq.3), taking into consideration the arbitrary beam shape, additional contributions of the Legendre functions presented in [25] are required, which might cause the data processing to be less effective, in terms of duration of processing, compared to the simple case of the Mie calculation presented before. In this section, we will provide a numerical approach for the above-mentioned scattering problem of a rotated incident plane wave, where the solution will be given in terms of the unrotated global system. The analytical expression is similar to the basic case of Mie scattering without any use of the beam shape coefficients, required in the GLMT. In addition, by using the error estimation technique, we can estimate the impact of the sampling densities, which will be shown to play a significant roles in the case of large tilting angles.

Fig. 3 resembles the scenario, presented in Fig. 1, while the incident wave comes with an angle. In order to describe the scattering problem in a global coordinate system, three Euler angles $\alpha, \beta, \gamma$ are defined. The global system $r_o$ is the one, considered in Fig. 1. The coordinates *x',y',z'* form an intermediate system, accounting for rotations $\alpha, \beta$ [Fig.3 (a)]. The system *x'',y'',z''*, denoted $\tilde{r}_o$, is the rotated system resulted by the rotations $\alpha, \beta$ and the tilt $\gamma$ along z' due to the new polarization orientation. The propagation direction $\hat{k}_i$ is oriented along *z''*, where $E_i$ and $B_i$ are along *x''* and *y''*, respectively.

In Fig.3 (b), similar to the previous section, the scattered field is described by spherical coordinates that can be related to the global system $r_o$ by $[r, \theta, \phi]$; or can be given in terms of $\tilde{r}_o$ by $[r, \tilde{\theta}, \tilde{\phi}]$.

As it was mentioned before, the solution in $r_o$ system is more complicated than $\tilde{r}_o$, coordinates, where the later is the Mie scattering case shown in the previous section. Therefore, in this section, we will relate the mathematical expression to the global system $r_o$, while maintaining the simplicity provided by solving the problem relatively to the rotated system $\tilde{r}_o$. In other words, the final solution will be given in terms of $r_o$, but the calculations will be based on $\tilde{r}_o$. In order to do so, the variables $\tilde{\theta}, \tilde{\phi}$ and $\theta, \phi$ will be associated by the Euler angles $\alpha, \beta, \gamma$.

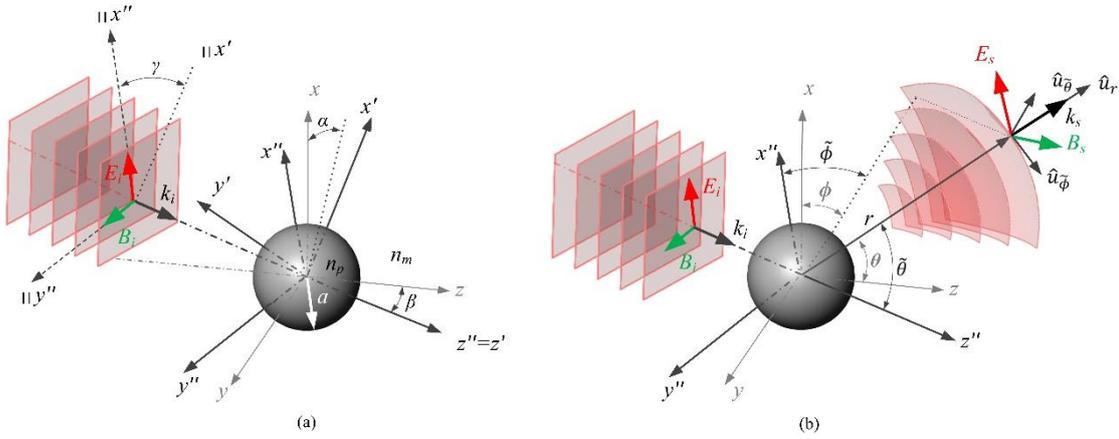

Fig. 3. Geometry of the scattering problem with a tilted incoming plane wave. (a) the incoming plane wave that propagates along the intermediate and the rotated systems with associated Euler angles; (b) the relation between the spherical coordinates, the unrotated system $r_0$ and the rotated system $\tilde{r}_0$.

The unit vector $\hat{u}_{r_0}(\theta, \phi) = sin\theta cos\phi \hat{u}_x + sin\theta sin\phi \hat{u}_y + cos\theta \hat{u}_z$ represents an arbitrary direction in terms of $r_o$. The transformation matrix $\tilde{A}$ (is given in the appendix) provides the expression of vector $\hat{u}_{r_0}(\theta, \phi)$ in terms of $r''_o$ as follows:

$$\hat{u}_{\tilde{r}_0}(\theta, \phi) = \tilde{A}(\alpha, \beta, \gamma) \cdot \hat{u}_{r_0}(\theta, \phi). \tag{9}$$

Next, the angles $\tilde{\theta}, \tilde{\phi}$, can be expressed by the following geometrical relations:

$$\tilde{\theta}(\theta, \phi; \alpha, \beta, \gamma) = \text{acos}(\hat{u}_{\tilde{r}_0}(\theta, \phi) \cdot \hat{z}''), \tag{10}$$

$$\tilde{\phi}(\theta, \phi; \alpha, \beta, \gamma) = \text{atan}\left(\frac{\hat{u}_{\tilde{r}_0}(\theta,\phi)\cdot \hat{y}''}{\hat{u}_{\tilde{r}_0}(\theta,\phi)\cdot \hat{x}''}\right). \tag{11}$$

Following Eqs. (9-11), the scattering amplitude can now be calculated in the global coordinate system.

In order to underline the nonlinear nature of the transformations (Eqs. 10 and 11), a special case of rotation along $x$ axis will be considered. Figure 4 (a-c) are the colormaps of $\tilde{\theta}(\theta, \phi)$ for $\beta = 0, \frac{\pi}{12}, \frac{\pi}{6}$, respectively, where $\alpha =$

$\gamma = 0$. Figure 4 (d, e) show two cross sections of the transformations, along the vertical planes $\phi = \frac{\pi}{2}$ and $\phi = 1.9\pi$, and clearly show the nonlinearity on the transformation.

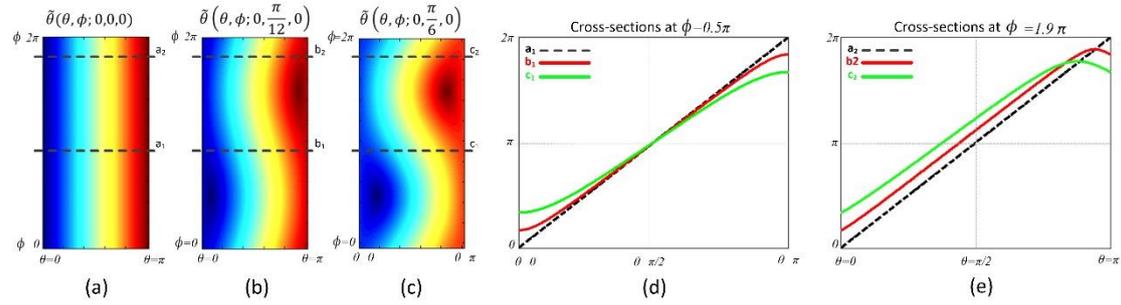

Fig. 4. Color map of $\tilde{\theta}(\theta, \phi)$. (a) the incident wave is parallel to the global system; (b,c) the incident wave is tilted in $\frac{\pi}{12}$ and $\frac{\pi}{6}$ along *x*, respectively. (d,e) two cross-sections along $0.5\pi$ and $1.9\pi$ (denoted $a_{1,2}$, $b_{1,2}$, and $c_{1,2}$, respectively), corresponded to the distributions presented in (a-c).

Figure 5 shows three scattering amplitudes (normalized, in logarithmic scale) for different incident angles, chosen to present the data in Fig. 4. The particle's size parameter is $k_m a = 5$ ($n_p$=1.5, $n_m$=1.33). According to the results (presented in Fig.5) one cane realized the rotation of the scattering amplitude with the incident angle, as expected, verifying the of the applied set of transformations.

According to Fig.5, one can realize that with the increasing in the angle $\beta$, the corresponded scattering amplitude distributions are changed in both orthogonal planes $\phi = 0°$ and $\phi = 90°$ (i.e. *z-x* and *z-y* planes), represented by the green and red lines, respectively. Additionally, in $\phi = 90°$, the symmetry breaks, where the forward scattering directions is tilted according to the angle $\beta$.. Nonlinearity of angular (Eqs 10 and 11) affects the sampling distributions $\Delta\tilde{\theta}$ and $\Delta\tilde{\phi}$, which are no longer uniform. However, using the error estimation presented in the previous section, we can straightforwardly calculate the error with respect to the change in the Euler angles. Following Eq. (8), we calculated $\eta_{scat}$ with respect to different rotations along the *x* axis, i.e. $\alpha = \gamma = 0$, where $\beta$ is varied. In addition, the sampling density rate is *L*=64,128,256, and 512.

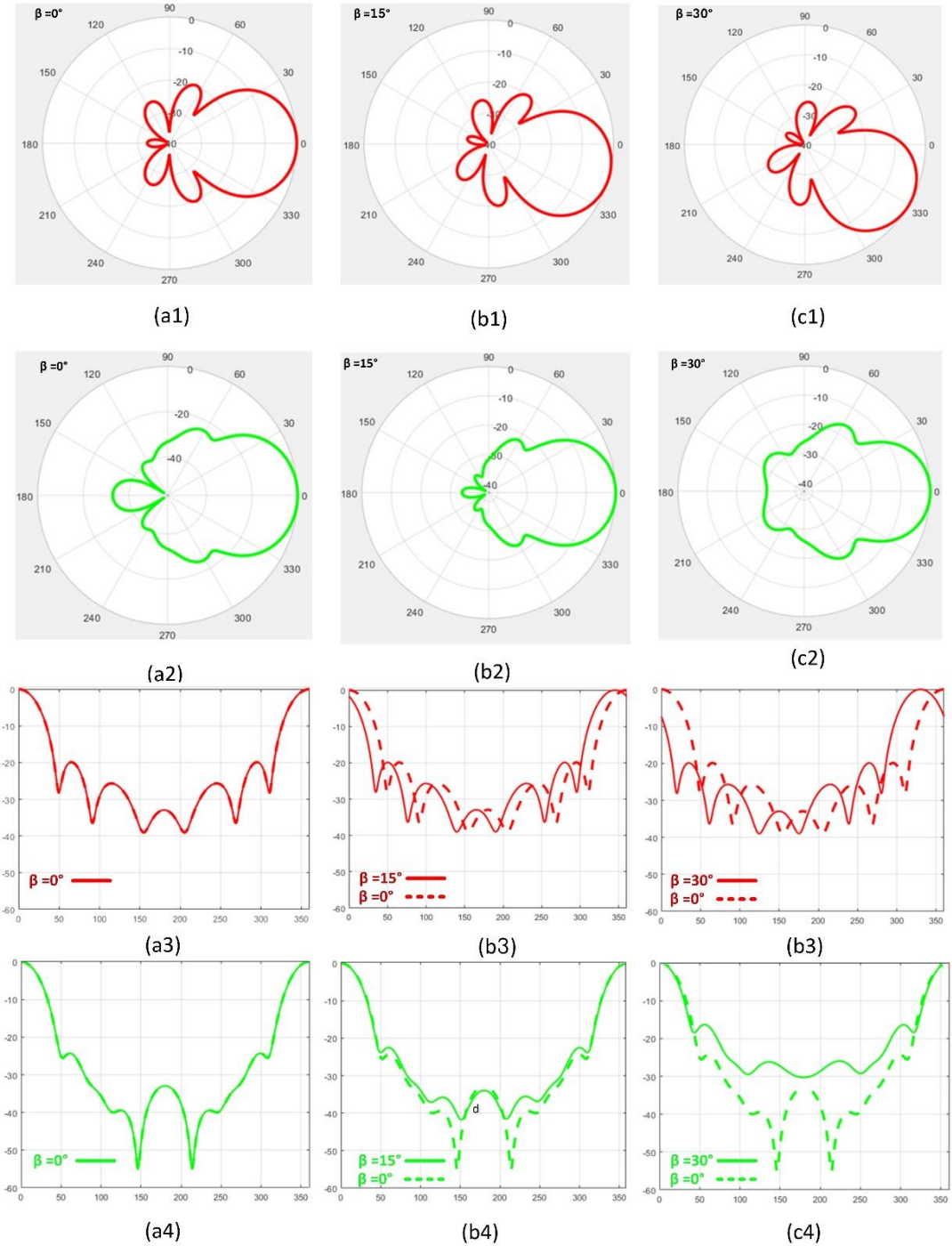

Fig.5 Plots of the scattering amplitude versus the angle $\theta$ for three different incident angles $\beta = 0, \frac{\pi}{12}, \frac{\pi}{6}$: (a1-c1) polar representation along H-plane; (a2-c2) polar representation along E-plane; (a3-c3) two dimensional plot along H-plane; (a4=c4) two dimensional plot along E-plane.

Fig.6 (a) shows the numerical error for several sampling densities and underlines the impact of the transformation on the accuracy. It can be seen that the numerical error grows with increasing the angle of incidence in respect to the optical axis (Fig. 1) and become quite significant in cases where extremely high numerical apertures are required. Furthermore, increasing the size of the particles leads to appearance of fast-oscillating features, which can be distorted by the rotation. Fig. 6(b) demonstrates this effect by analyzing the numerical arrow for four different four size parameters $k_m a = 0.1, 2.5, 5$ and $10$ (the sampling density rate here is $L$=512).

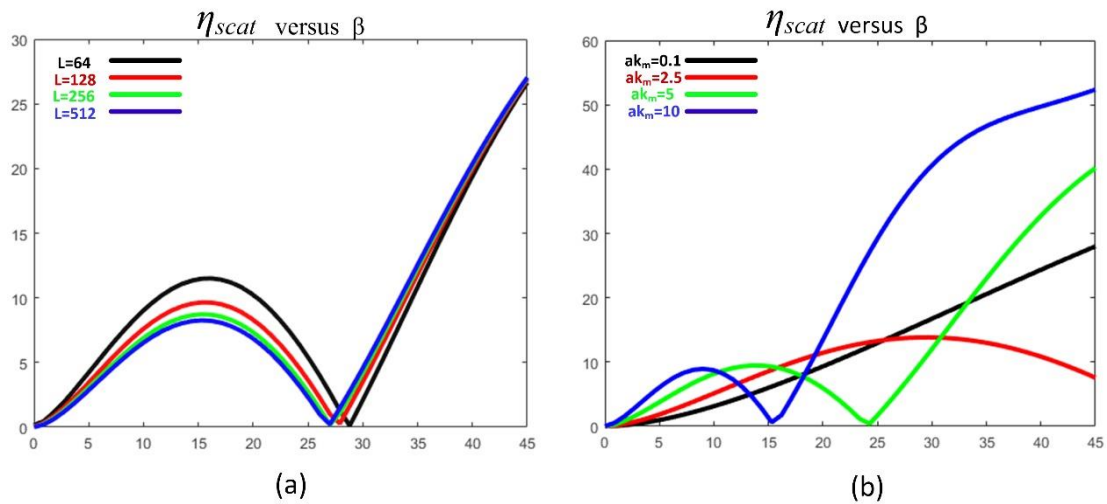

(a)  (b)

Fig.6 The numerical error resulted by rotations along the *x* axis. (a) the numerical error resulted by rotations along the *x* axis according to four different sampling density rates; (b) the numerical error according to four different size parameters for $\beta = [0,45°]$

## 4. Scattering of an arbitrary incident waveform

In this part we will present an efficient numerical approach to calculate the scattering for an arbitrary wave front. The illustration shown in Fig. 7, describes a special case of a focused linearly polarized Gaussian beam.

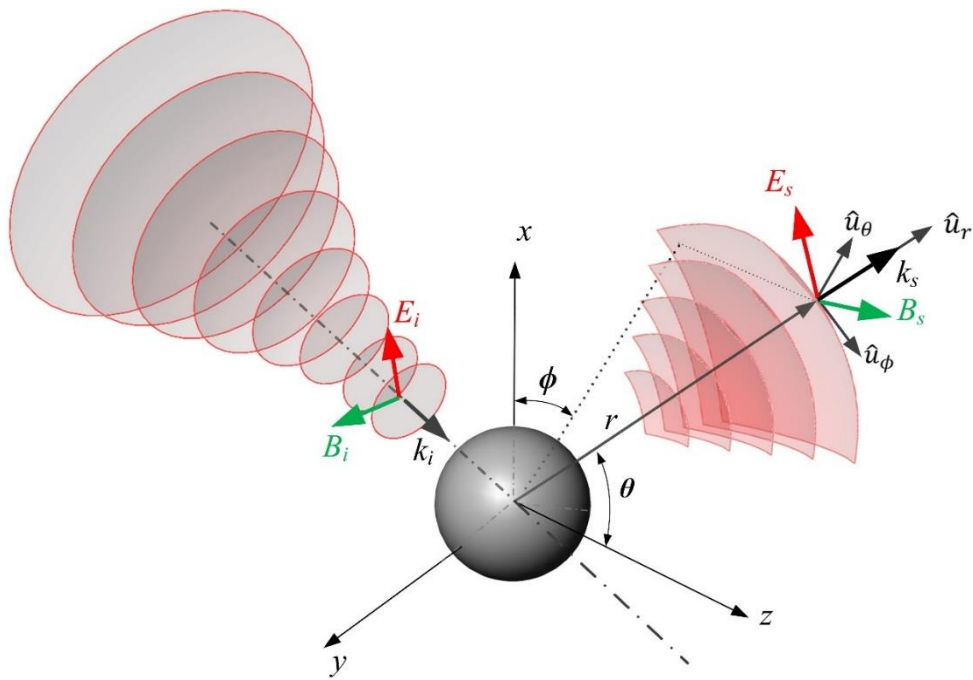

Fig. 7. Geometry of a general scattering problem.

Incident beam in the spatial frequency domain (also known as the $k$ space) is represented as follows:

$$\widetilde{\boldsymbol{E}}_i(\boldsymbol{k}_t) = \int \boldsymbol{E}_i(\boldsymbol{r}_t) exp(i\boldsymbol{k}_t \cdot \boldsymbol{r}_t) d^2 r_t, \quad (12)$$

where, $\boldsymbol{r}_t = x\hat{x} + y\hat{y}$ is the transverse space located at the center of the particle; $\boldsymbol{k}_t = k_x\hat{x} + k_y\hat{y}$ is the transverse spectral wavenumber corresponded to $\boldsymbol{r}_t$, such that $\boldsymbol{k}_t \cdot \boldsymbol{r}_t = n_m(k_x x + k_y y)$; the tilde sign (~) denotes a Fourier transformed quantity. The resulted $\widetilde{\boldsymbol{E}}_i(\boldsymbol{k}_t)$ consists of multiple plane waves, each of which is oriented and linearly polarized in different direction and can have an additional phase factor. Three Euler angles (shown in Fig. 3, section 3) are needed for a complete representation. Following the plane wave decomposition method presented in [27], any given plane wave can be represented by a superposition of Transverse-Electric (TE) and Transverse-Magnetic (TM) components (represented by $\widetilde{E}_i^{(TE)}$ and $\widetilde{E}_i^{(TM)}$, respectively) as follows:

$$\widetilde{\boldsymbol{E}}_i(\boldsymbol{k}_t) = \widetilde{E}_i^{(TE)} \hat{\boldsymbol{n}}(\boldsymbol{k}_t) + \widetilde{E}_i^{(TM)} \hat{\boldsymbol{t}}(\boldsymbol{k}_t). \quad (13)$$

The unit vectors $\hat{\boldsymbol{n}}(\boldsymbol{k}_t)$ and $\hat{\boldsymbol{t}}(\boldsymbol{k}_t)$ represent the normal and transversal directions, relatively to the plane of incident:

$$\hat{\boldsymbol{n}}(\boldsymbol{k}_t) = \frac{1}{k_t}(k_y \hat{x} - k_x \hat{y}), \quad \hat{\boldsymbol{t}}(\boldsymbol{k}_t) = \frac{k_z}{k_m k_t}(k_x \hat{x} + k_y \hat{y}) - \frac{k_t}{k_m}\hat{z}, \quad (14)$$

where, $k_t = \sqrt{k_x^2 + k_y^2}$ and $k_z = \sqrt{k_m^2 - (k_x^2 + k_y^2)}$.

The magnitudes $\widetilde{E}_i^{(TE)}$ and $\widetilde{E}_i^{(TM)}$ can be calculated by a scalar multiplication as follows:

$$\begin{aligned}\widetilde{E}_i^{(TE)} &= \widetilde{\boldsymbol{E}}_i(\boldsymbol{k}_t) \cdot \hat{\boldsymbol{n}}(\boldsymbol{k}_t) = \frac{1}{k_t}\left(k_y \widetilde{E}_{i_x} - k_x \widetilde{E}_{i_y}\right) \\ \widetilde{E}_i^{(TM)} &= \widetilde{\boldsymbol{E}}_i(\boldsymbol{k}_t) \cdot \hat{\boldsymbol{t}}(\boldsymbol{k}_t) = \frac{1}{k_z k_t}\left(k_x \widetilde{E}_{i_x} + k_y \widetilde{E}_{i_y}\right).\end{aligned} \quad (15)$$

Once the TE-TM ratio and the propagation orientation in the spatial frequency domain are known (Eq.15), three Euler angles, corresponding to $\boldsymbol{k}_t$ can extracted as follows:

$$\begin{aligned}\alpha &= acos\left(\frac{k_z}{k_m}\right) \\ \beta &= acos\left(\frac{k_x}{k_t}\right) \\ \gamma &= atan\left(-\frac{\widetilde{E}_i^{(TM)}}{\widetilde{E}_i^{(TE)}}\right)\end{aligned} \quad (16)$$

According to Eq. 10 and 11, presented in section 3, the warped angles $\widetilde{\theta}$ and $\widetilde{\phi}$ can now be determined for every $k_x, k_y$, according to Eq.16.

Next, according to Eqs.2, 3 (presented in section 2), for any incoming plane wave (in the spatial frequency domain) the resulted scattered electric and magnetic fields $\widetilde{\boldsymbol{E}}_s(\boldsymbol{k}_t)$, $\widetilde{\boldsymbol{B}}_s(\boldsymbol{k}_t)$ can be calculated in respect to the input argument $\widetilde{\theta}(\theta, \phi; \alpha(\boldsymbol{k}_t), \beta(\boldsymbol{k}_t), \gamma(\boldsymbol{k}_t))$ and $\widetilde{\phi}(\theta, \phi; \alpha(\boldsymbol{k}_t), \beta(\boldsymbol{k}_t), \gamma(\boldsymbol{k}_t))$.

Finally, applying inverse Fourier transform enables to derive the scattered fields of the arbitrary incident wave front, denoted $\boldsymbol{E}_s^{tot}$ and $\boldsymbol{B}_s^{tot}$, respectively:

$$\boldsymbol{E}_s^{tot}(\boldsymbol{r}_0) = \frac{e^{jk_m r}}{(2\pi)^2 r} \int \widetilde{\boldsymbol{E}}_i(\boldsymbol{k}_t) \boldsymbol{f}\left(\widehat{\boldsymbol{k}}_s(\tilde{\theta},\tilde{\phi}), \widehat{\boldsymbol{k}}_i(\tilde{\theta},\tilde{\phi})\right) exp(-j\boldsymbol{k}_t \cdot \boldsymbol{r}_0) d^2 k_t k_t,$$

$$\boldsymbol{B}_s^{tot}(\boldsymbol{r}_0) = \frac{e^{jk_m r}}{c(2\pi)^2 r} \int \widehat{\boldsymbol{k}}_s(\tilde{\theta},\tilde{\phi}) \times \widetilde{\boldsymbol{E}}_i(\boldsymbol{k}_t) \boldsymbol{f}\left(\widehat{\boldsymbol{k}}_s(\tilde{\theta},\tilde{\phi}), \widehat{\boldsymbol{k}}_i(\tilde{\theta},\tilde{\phi})\right) exp(-j\boldsymbol{k}_t \cdot \boldsymbol{r}_0) d^2 k_t k_t.$$
(17)

To implement a numerical calculation, the space $\boldsymbol{k}_t$ is discretized by $k_1 = -K_1,..,0,...K_1$ and $k_2 = -K_2,..,0,...K_2$, where $K_1, K_2 \leq k_m$ represent the maximal frequency along the horizontal and vertical directions in the spatial frequency domain, respectively. The FT is implemented by a Discrete Fourier Transform (DFT), for example [28], where the scaling factor (dependent by $\lambda_0$ and $n_m$) is taken into consideration. Once the parameter size and the medium are known, a set of resulted scattering amplitudes, denoted $\bar{\boldsymbol{f}}(\theta_{l_1},\phi_{l_2};k_1,k_2)$, corresponded to the arbitrary $\widetilde{\boldsymbol{E}}_i(k_1,k_2)$ can be calculated and saved to a generalized scattering amplitudes database. Next, according to the inverse FT presented in Eq.17, the resulted scattered fields can be integrated as follows:

$$\boldsymbol{E}_s^{tot}(\boldsymbol{r}_0) = \frac{exp(-jk_m r)}{(2\pi)^2 r} \sum_{k_1,k_2} \widetilde{\boldsymbol{E}}_i((k_1,k_2)) \bar{\boldsymbol{f}}(\theta_{l_1},\phi_{l_2};k_1,k_2) \Delta k_1 \Delta k_2$$

$$\boldsymbol{B}_s^{tot}(\boldsymbol{r}_0) = \frac{exp(-jk_m r)}{c(2\pi)^2 r} \sum_{k_1,k_2} \widehat{\boldsymbol{k}}_s(\theta_{l_1},\phi_{l_2}) \times \widetilde{\boldsymbol{E}}_i((k_1,k_2)) \bar{\boldsymbol{f}}(\theta_{l_1},\phi_{l_2};k_1,k_2) \Delta k_1 \Delta k_2,$$
(18)

where, $\Delta k_1 = 2K_1/M$ and $\Delta k_2 = 2K_2/N$ is the sampling density at the spatial domain. M and N represent the amount of pixels along the horizontal and vertical dimensions, respectively.

By doing so, for every arbitrary phase pattern, the DFT can be calculated and multiplied by the generic scattering amplitude $\bar{\boldsymbol{f}}(\theta_{l_1},\phi_{l_2};k_1,k_2)$, which is not dependent by the incident phase pattern or its polarization. In other words, the scattering amplitude, which is dependent by the arguments $k_x, k_y$ can be retrieved from a scattering amplitude database that was calculated in advance and is needed to be calculated only once.

The example, presented in Fig. 9 demonstrates the above mentioned special case of a focused Gaussian beam linearly polarized along *x* direction, with a wavelength of $\lambda_0 = 1\ \mu m$ and three different cases of beam waists, $w_0 = 1, 5$ and $10\ \mu m$. The discrete space consists of N=M=65 pixels with a pixel size of $\lambda_0/25$. The refractive index of the sphere is $n_p$=1.5 and it is surrounded by air ($n_m$=1). Moreover, the corresponded spatial frequency (*k*-space) is limited by $K_1 = K_2 = 0.05 k_m$. In other words, the maximal $\beta$ angle involved in the integration is $\beta_{max} = 4.05°$.

The duration of time required to retrieve Mie solutions from the data base and integration (using Matlab) was 2-5 seconds.

From Fig.9 (a-d), one can realize that a beam waist, which is 20 times larger than the sphere's radius, is resulting in a scattering amplitude (represented by the solid line) that is identical to the scattering amplitude corresponded to an equivalent plane wave

(represented by the dashed line) that is based on the same physical conditions (e.g. wavelength, refractive index of the medium). However, the smaller the waist (i.e. closer to the size of the spherical scatterer) additional spectral frequencies are becoming increasingly dominant.

In the previous sections we presented a technique to estimate the numerical error resulted by the limited sampling density. In particular, in section 3 we have analyzed the effect occurring due to the change in the field orientation relatively to the global system. The scattering error for different angles of incidence $\eta_{scat}(k_x, k_y)$ In Fig. 10, one can realize that the maximal error occurs in the maximal spectral frequencies, where $k_1 = K_2 = 0.05k$ (corresponding to $\beta_{max} = 4.05°$ ) is approximately 1.4%.

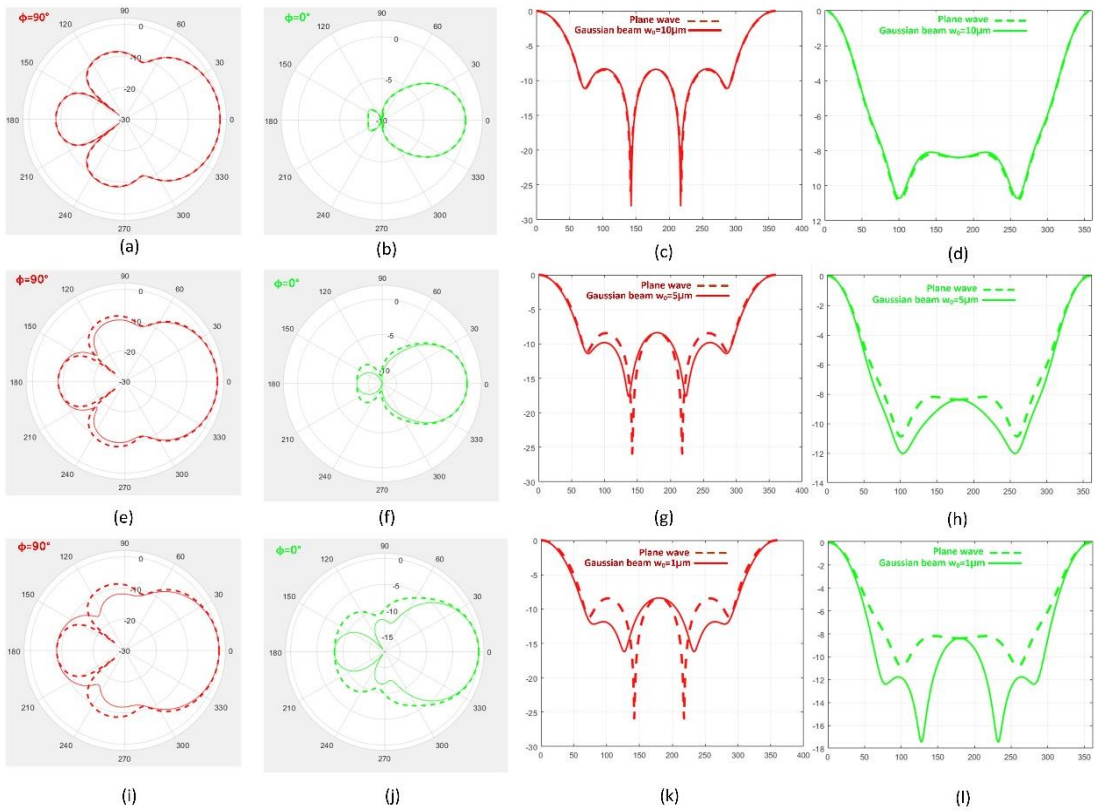

Fig.9. Scattering of a Gaussian beam on a dielectric sphere the scattering amplitude as a function of $\theta$. polar representation at the E and H plane and their two dimensional corresponded plots. (a-d) waist radius of 10 µm ; (e-h) waist radius of 5 µm ; (i-l) waist radius of 1 µm

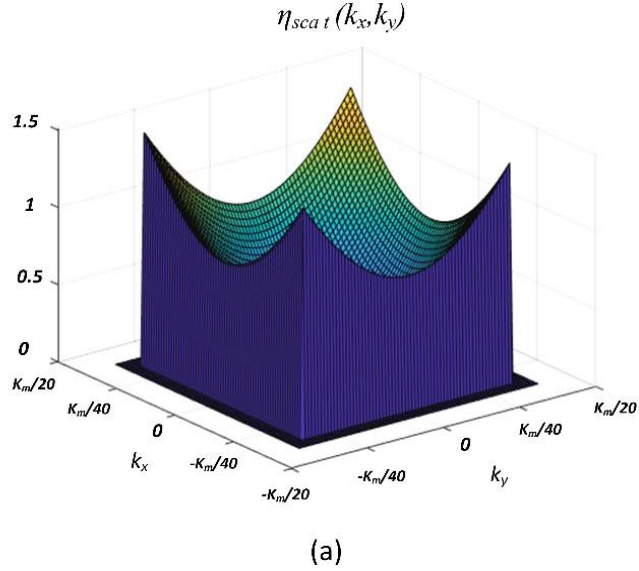

Fig. 10 scattering error; (a) tree dimensional mesh of $\eta_{scat}$ versus $k_x$ and $k_y$

## 5. Outlook and Conclusions

A problem of arbitrary wave front scattering from a sphere have been considered. Unlike the GLMT, where the scattering is calculated by the beam shape coefficient for each phase pattern (including tilted plane wave) in a different modified coordinate system, our approach is based on retrieving the data form an existing data base of Mie solutions. We perform a plane wave decomposition of the incident illuminations and apply a set of rotations to bring a weighted sum of Mie solutions (each one corresponds to a certain incident plane wave with its own phase and amplitude) to a global coordinate system. This method provides a significant computational advantage to address any incident waveform in a fast and efficient way.

A natural use case for the proposed technique is when real-time calculations of scattering problems are needed. Possible applications include scanning microscopy with focused beams [29], dynamic structured illumination-based imaging systems [30] and many others. Another example is holographic optical tweezers [31], where real-time arbitrary phase patterns are used to control particle's motion. Systems with active control require obtaining real-time feedback [32], based on solution of scattering problem from a moving particle [33], [34], the electromagnetic forces and is based on the scattered field, can be easily calculated and updated in real time.

## Appendix A

In general, the rotation matrix $\tilde{A}(\alpha,\beta,\gamma)$ can be represented by the following nine components as follows

$$\tilde{A}(\alpha,\beta,\gamma) = \begin{pmatrix} \tilde{A}_{11} & \tilde{A}_{12} & \tilde{A}_{13} \\ \tilde{A}_{21} & \tilde{A}_{22} & \tilde{A}_{23} \\ \tilde{A}_{31} & \tilde{A}_{32} & \tilde{A}_{33} \end{pmatrix}, \tag{A1}$$

where each of the component is given by the Euler angles

$$\begin{aligned}
\tilde{A}_{11} &= cos(\gamma)cos(\beta) \\
\tilde{A}_{12} &= sin(\gamma)sin(\alpha)cos(\beta) - cos(\gamma)sin(\beta) \\
\tilde{A}_{13} &= cos(\gamma)sin(\alpha)cos(\beta) + sin(\gamma)sin(\beta) \\
\tilde{A}_{21} &= cos(\alpha)sin(\beta) \\
\tilde{A}_{22} &= sin(\gamma)sin(\alpha)sin(\beta) + cos(\gamma)cos(\beta) \\
\tilde{A}_{23} &= cos(\gamma)sin(\alpha)sin(\beta) - sin(\gamma)cos(\beta) \\
\tilde{A}_{31} &= -sin(\alpha) \\
\tilde{A}_{32} &= sin(\gamma)cos(\alpha) \\
\tilde{A}_{33} &= cos(\gamma)cos(\alpha)
\end{aligned} \tag{A2}$$